\documentclass[prb,aps,twocolumn,draft,tightenlines, %
showpacs,floatfix]{revtex4}
\usepackage{psfig}
\usepackage{amssymb}
\usepackage{amsmath}

\begin{document}
\title{Control of spin coherence in $n$-type GaAs quantum wells
using strain}
\author{L. Jiang}
\author{M. W. Wu}
\thanks{Author to whom all correspondence should be addressed}%
\email{mwwu@ustc.edu.cn.}
\affiliation{Hefei National Laboratory for Physical Sciences at
  Microscale,University of Science and Technology of China, \\ and
  Department of Physics,University of Science and Technology of China,
  Hefei, Anhui, 230026, China}
\altaffiliation{Mailing Address.}
\date{\today}

\begin{abstract}
We show that the bulk-inversion-asymmetry-type strain-induced 
spin-orbit coupling can be used to effectively modify the
Dresselhaus spin splitting in (001) GaAs quantum wells
with small well width and the resulting spin dephasing time can be
increased by two orders of magnitude to nanoseconds
under right conditions. 
The efficiency of this strain manipulation of the
spin dephasing time under different conditions such as
temperature, electric field and electron density is
investigated in detail.
\end{abstract}
\maketitle

Manipulation of the spin coherence/dephasing in Zinc-blend semiconductors,
where the symmetry of the spin degrees of freedom is
broken due to the lack of inversion center of the crystal,
is one of the fundamental
subjects in semiconductor spintronics\cite{wolf,spintronics,das}
 which aims to incorporate the
spin degrees of freedom into the traditional electronic devices.
It has been shown both experimentally and 
  theoretically that many effects, such as magnetic field and electric
  field, can strongly affect the spin precession and  
spin  dephasing.\cite{kikka,sih,loss,Wu, weng, multi} 
Very recently strain has also been shown to be effective in spin manipulation.
Kato {\em et al.}
reported experimentally that strained bulk semiconductors 
exhibit spin splitting in the
presence of applied electric fields.\cite{kato1}
They further used this strain-induced spin splitting 
to generate spin polarization in the presence of 
an electric current.\cite{kato2} 
In this report we demonstrate that strain can also be used to effectively 
control the spin coherence and greatly enhance the spin dephasing time (SDT). 

The leading spin dephasing mechanism in $n$-type GaAs quantum well (QW)
in the absence of applied electric field along the growth direction
is the D'yakonov-Perel' mechanism\cite{dp} due to the   
 Dresselhaus\cite{dres} spin splitting ${\bf h}({\bf k})
\cdot\mbox{\boldmath$\sigma$\unboldmath}/2$.\cite{Jusserand, Wayne} 
In (001) QW with the growth direction along the
$z$-axis, ${\bf h}({\bf k})$ contains terms both linear and
cubic in $k$.  When only the lowest subband 
is populated, it reads  $h_x(\mathbf{k}) 
  = -\gamma k_x (\langle k_z^2\rangle-k^2_y)$, $h_y(\mathbf{k}) =
  \gamma k_y ({\langle k_z^2\rangle}-k_x^2)$ and $h_z({\bf k})=0$
with $\gamma$ denoting the
spin-orbit coupling strength\cite{Aronov} and $\langle
  k_z^2\rangle$ representing the average of the operator
  $-(\partial/\partial z)^2$ over the electronic state of the lowest
  subband. Under the infinite-well-depth assumption, 
$\langle k_z^2\rangle=(\frac{\pi}{a})^2$ with $a$ standing for the well width.
It has been shown very
recently\cite{multi} from a full 
many-body kinetic study  
of the spin dephasing that for narrow well width 
with $\pi^2/a^2\gg\langle k_x^2\rangle$ and  
$\pi^2/a^2\gg\langle k_y^2\rangle$, 
the linear term in ${\bf h}({\bf k})$ is dominant and the SDT {\em increases}
 with temperature. 
Here $\langle\cdots\rangle$ stands for the average subject to the
Fermi distribution. However, when the 
well width is big enough and/or the temperature is high enough that
the cubic term is dominant, the SDT decreases with temperature as 
commonly expected.\cite{multi} 

Strain introduces additional spin splittings
and the leading one is the one of bulk-inversion-asymmetry type:\cite{Pikus}
$h^{s}_x(\mathbf{k}) = -D k_x (\epsilon_{yy} - \epsilon_{zz})$, 
$h^{s}_y(\mathbf{k}) = -D k_y (\epsilon_{zz} - \epsilon_{xx})$,
$h^{s}_z(\mathbf{k}) = -D k_z (\epsilon_{xx} - \epsilon_{yy})$,
and is linear in $k$. $D$ is the material constant.
 Therefore the linear term of the Dresselhaus spin
splitting ${\bf h}({\bf k})$ can be adjusted by the strain and the 
total spin splitting term can be written as ${\bf h}^t({\bf k})\cdot
\mbox{\boldmath$\sigma$\unboldmath}/2$ with
\begin{eqnarray}
\label{htx}
&&h^t_x(\mathbf{k})= [(-\alpha + \beta) + \gamma k_y^2] k_x\ ,\\
\label{hty}
&& h^t_y(\mathbf{k})=  -[(-\alpha + \beta) + \gamma k_x^2] k_y\ ,
\end{eqnarray}
and $h^t_z(\mathbf{k})=0$
by taking the strain $\epsilon_{xx} = \epsilon_{yy}$ and
$\epsilon_{zz} - \epsilon_{xx} > 0$.\cite{kato1,kato2} 
 $\alpha=\gamma(\pi/a)^2$  and $\beta=D\epsilon$ 
with $\epsilon=\epsilon_{zz}-\epsilon_{xx}$. 
 Equations\ (\ref{htx}) and (\ref{hty}) clearly indicate that 
under certain well width and  strain, $\alpha=\beta$ and the 
spin splitting can be totally determined by the cubic term. 
In addition, by modulating the magnitude of the strain, the
relative magnitudes of the linear and cubic terms
are varied. Different dependences of the SDT on the external conditions
such as temperature, electric field and electron density
are therefore expected under different strains.
Finally one may dramatically suppress the spin dephasing
by adjusting the strain to satisfy the condition 
$\alpha-\beta=\gamma\langle k_\xi^2\rangle$ with $\xi=x$, $y$.

We construct the many-body kinetic spin Bloch equations \cite{Wu1}
by the non-equilibrium Green function method\cite{Haug}
as follows:\cite{weng}
 \begin{equation}
\label{BEQ}
 \dot{\rho}_{\mathbf{k}, \sigma \sigma^{\prime}}
 -e\mathbf{E}\cdot\mathbf{\bigtriangledown}_{\mathbf{k}}\rho_{\mathbf{k},
 \sigma \sigma^{\prime}}=\dot{\rho}_{\mathbf{k}, \sigma
 \sigma^{\prime}}|_\mathtt{coh}+\dot{\rho}_{\mathbf{k}, \sigma
 \sigma^{\prime}}|_\mathtt{scatt} 
  \end{equation}
with $\rho_{\mathbf{k}, \sigma \sigma^{\prime}}$ representing the 
single-particle density matrix elements. 
The diagonal elements $\rho_{\mathbf{k}, \sigma \sigma}\equiv
f_{\mathbf{k}, \sigma}$  describe the
 electron distribution 
functions  of wavevector $\mathbf{k}$ and
spin $\sigma$ ($= \pm 1/2$). The
  off-diagonal elements $\rho_{\mathbf{k}, \frac{1}{2} -\frac{1}{2}} =
  \rho_{\mathbf{k}, -\frac{1}{2} \frac{1}{2}}^{\ast} \equiv
  \rho_{\mathbf{k}}$ describe the  inter-spin-band correlations 
  for the spin coherence. The second term in the kinetic
equations describes the momentum and energy input from a
uniform  external electric field $\mathbf{E}$ along the
$x$-axis. $\dot{\rho}_{\mathbf{k}, \sigma
    \sigma^{\prime}}|_\mathtt{coh}$ on the right hand side of the
  equations describes the coherent spin precession around the applied
  magnetic field $\mathbf{B}$ (along the $x$-axis, {\em i.e.}, in
the Voigt configuration), the effective magnetic field
  $\mathbf{h}^t(\mathbf{k})$  as well as the effective
  magnetic field from the electron-electron interaction in the
  Hartree-Fock approximation:
\begin{widetext}
  \begin{equation}
    \label{eq:f_coh}
\left. \frac{\partial f_{{\bf k},\sigma}} {\partial t}
    \right|_{\mathtt{coh}}=
    -2\sigma\bigl\{[g\mu_BB+h^t_x({\bf k})]\mbox{Im}\rho_{{\bf k}}+h^t_y({\bf k})
    \mbox{Re}\rho_{{\bf k}}\bigr\}
    +4\sigma\mbox{Im}\sum_{{\bf q}}V_{{\bf q}}\rho^{\ast}_{{\bf k}+{\bf
        q}} \rho_{{\bf k}},
  \end{equation}
  \begin{eqnarray}
    \label{eq:rho_coh}
  \left.\frac{\partial \rho_{{\bf k}}}{\partial t}
      \right|_{\mathtt{coh}} &=&
      \frac{1}{2}[ig\mu_B B + ih^t_x({\bf k}) + h^t_y({\bf k})]
      (f_{{\bf k}\frac{1}{2}}-f_{{\bf k}-\frac{1}{2}})\nonumber\\
      &&+i\sum_{{\bf q}}V_{\bf q}\bigl[(f_{{\bf k}+{\bf q}\frac{1}{2}}
      -f_{{\bf k}+{\bf q}-\frac{1}{2}})\rho_{{\bf k}}
      -\rho_{{\bf k}+{\bf q}}(f_{{\bf k}\frac{1}{2}}
      -f_{{\bf k}-\frac{1}{2}})\bigr]\ .
\end{eqnarray}
\end {widetext}     
$\dot{\rho}_{\mathbf{k}, \sigma
  \sigma^{\prime}}|_\mathtt{scatt}$ denotes the 
electron-electron, electron-phonon and
electron-impurity scattering. The expressions of these terms can
be found in Ref.\ \onlinecite{weng}.  
One notices that all the unknowns appear
 in the scattering terms. 
Therefore the kinetic Bloch
equations (\ref{BEQ}) have to be solved self-consistently to obtain the
temporal evolution of the electron distribution functions
$f_{\mathbf{k}, \sigma}(t)$ and the spin 
coherence $\rho_{\mathbf{k}}(t)$. The details of
the calculation are laid out in Ref.\ \onlinecite{weng}.
The SDT is obtained by the slope of
 the envelop of the 
incoherently summed spin coherence
$\rho=\sum_{\mathbf{k}}|\rho_{\mathbf{k}}(t)|$.\cite{Wu1,Haug,kuhn}
It is understood that both true dissipation and the
interference of many k states may contribute to the decay.
The incoherent summation is therefore used to isolate the
irreversible decay from the decay caused by interference.\cite{Haug,kuhn}

We include the electron-longitudinal
optical phonon and the electron-electron
Coulomb scattering in the calculation. The impurity density is taken to be
zero throughout the paper.
The main results of our calculation are
summarized in Figs.\ 1 to 4. In the calculation 
the material parameters are listed in
Ref.\ \onlinecite{weng}. 
The width of the QW  is fixed to be $10$~nm.
The material constant $D$ is chosen to be $D=1.59\times10^{4}$\ m/s
  following the experiment.\cite{kato1}

\begin{figure}[htbp]
    \centerline{
      \psfig{figure=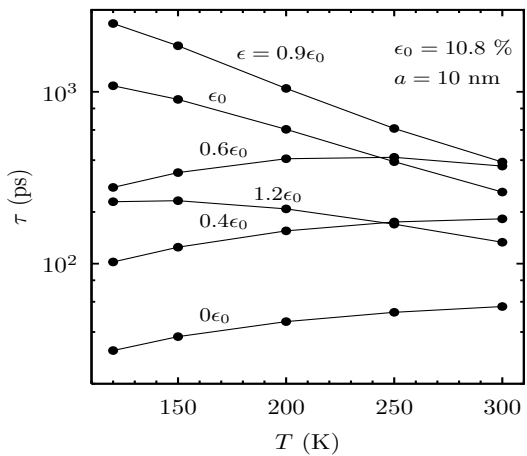,width=7cm,height=6cm}}
    \caption{SDT {\em vs.} the background temperature $T$
under different strains.  The electron density is
 $4\times10^{11}$~cm$^{-2}$.
} 
    \label{figjw1}
  \end{figure}

First we investigate the temperature dependence of the spin dephasing
under different strains. The SDT versus the background temperature
without an applied electric field is plotted in
Fig.\ 1. It shows that the temperature dependence of
the SDT under different strains is quite different. 
For small strain, say the strain is
smaller than $0.4\epsilon_0$  ($\epsilon_0\equiv\alpha/D$ denotes the strain
at which the linear term in ${\bf h}^t$ is exactly
eliminated), the linear term in ${\bf h}^t({\bf k})$ is
dominant and  the SDT {\em increases} monotonously  with the
temperature. For  strain around $\epsilon_0$, the contribution 
from the cubic term becomes important (or is the only
${\bf k}$-dependent term at $\epsilon=\epsilon_0$), the SDT 
either first increases then decreases with $T$ when there is
still linear term contribution or decreases with $T$ monotonically
when there is no linear term left ($\epsilon=\epsilon_0$).

These behaviors can be understood as follows:\cite{multi}
When the temperature increases, the electron-electron and
electron-phonon  scattering is enhanced. Consequently electrons
are driven to a more homogeneous state in ${\bf k}$-space. This tends
to increase the SDT. In the meantime, the increase of temperature also
drives electrons to a higher $k$-state and thus
induces a larger $h^t({\bf k})$.
This tends to reduce the SDT. Both linear and cubic terms of $h^t({\bf k})$ 
increase with $k$, but with a different increase rate. When the linear term
is dominant ({\em i.e.}, $|\alpha-\beta|>\gamma\langle k_\xi^2\rangle$), 
although its effect increases with temperature, 
the increase rate is slower than that of the scattering and 
the SDT increases with temperature. However, when the cubic term is dominant,
the effect of the cubic term increases much faster with temperature 
than the scattering and the SDT decreases with the temperature. This
effect is consistent with what obtained from  strain-free QW's.\cite{multi}

\begin{figure}[htbp]
  \centerline{
    \psfig{figure=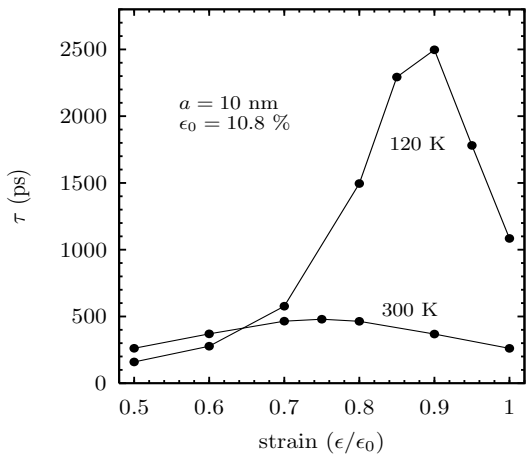,width=7cm,height=6cm}}
  \caption{SDT {\em vs.} strain at two temperatures. The electron
density is $4\times10^{11}$~cm$^{-2}$. 
}
    \label{figjw2}
  \end{figure}

From Fig.\ 1 one also notices that when the strain is applied, the SDT
can be greatly enhanced. At low temperature it can be as long as
{\em nanosecond} which is two orders of magnitude larger than the 
strain-free case.
In order to show the strain dependence of the SDT, we plot in Fig.\ 2 the
SDT as a function of strain for different temperatures. It is seen 
from the figure that the SDT first increases with strain 
until it reaches a maximum and then decreases with it. It is again noted that
at low temperature (120\ K) the varying range 
of the SDT versus 
$\epsilon$ sweeps over two orders of magnitude with the maximum 
SDT being 2.5\ ns. It is known that for QW with small width, the SDT is
in the order of tens of picoseconds. The present results 
indicate the possibility of using strain to obtain a very long
SDT in GaAs QW's.

\begin{figure}[htbp]
\vskip 0.5cm
  \centerline{
    \psfig{figure=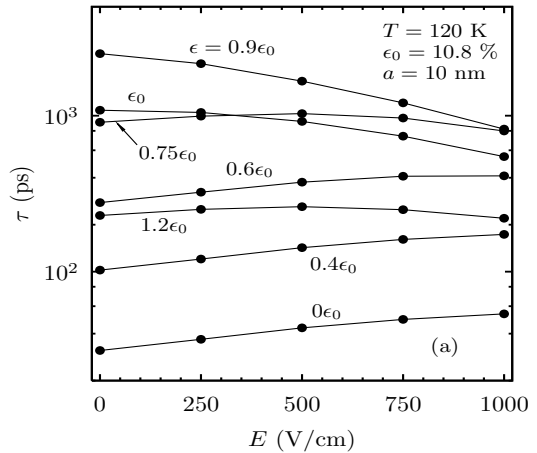,width=7cm,height=6cm}}
\vskip 0.5cm
  \centerline{
    \psfig{figure=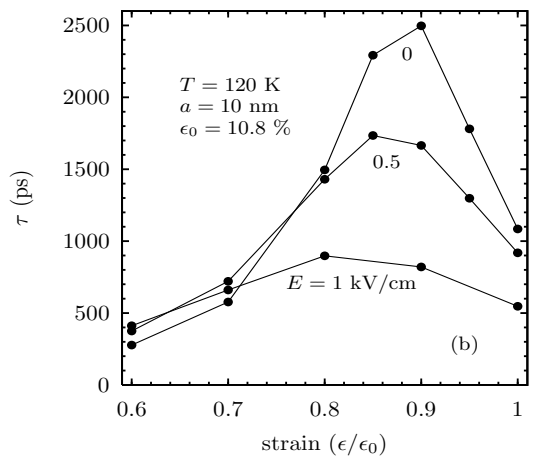,width=7cm,height=6cm}}
  \caption{Electron field dependence of the SDT.
SDT {\em vs.} the applied electric field $E$ under
different strains (a) and the strain $\epsilon$ at
different electric fields (b).
 The electron 
      density is $4\times10^{11}$~cm$^{-2}$. 
}
    \label{figjw3}
  \end{figure}

The physics of the $\tau$-$\epsilon$ dependence can be understood as following:
For QW with $a=10$\ nm, $\alpha=1.72\times 10^3$\ m/s. When $T=120$\ K and
electron density is 4$\times 10^{11}$\ cm$^{-2}$,
$\gamma\langle k_x^2\rangle=\gamma\langle k_y^2\rangle=
2.21\times 10^2$\ m/s. Therefore 
for strain-free case ($\beta=0$) the linear term in Eqs.\ (\ref{htx})
and (\ref{hty}) is one order of magnitude larger  than the cubic term.
Introducing a positive strain reduces the linear term, at certain 
strain  $\gamma\langle k_x^2\rangle-(\alpha-\beta)=0$ and
$h^t({\bf k})$ is greatly suppressed. Therefore one obtains 
a very large SDT.
$\epsilon$ predicted from above equation at 120\ K (300\ K) is 
$0.87\epsilon_0$ ($0.74\epsilon_0$ as $\gamma\langle k_x^2\rangle=
4.53\times10^2$\ m/s at 300\ K), which is in good agreement with 
$0.9\epsilon_0$ ($0.75\epsilon_0$) in Fig.\ 2.

\begin{figure}[htbp]
\vskip 0.5cm
  \centerline{
    \psfig{figure=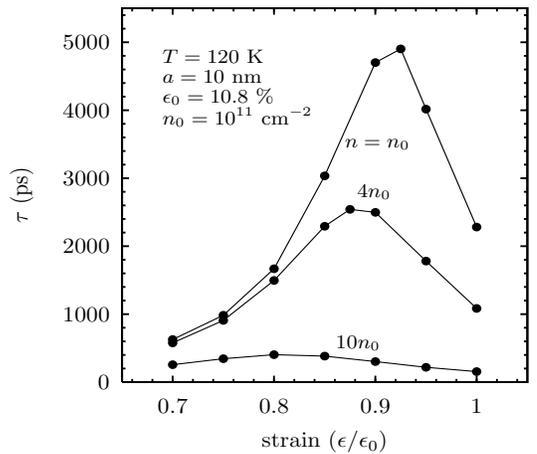,width=7cm,height=6cm}}
  \caption{SDT {\em vs.} the strains at different
    electron densities.
$T=120$\ K.}
   \label{figjw5}
  \end{figure}

Next we turn to the problem of  the applied-electric-field dependence of
the SDT under different strains. In Fig. 3(a), the SDT is plotted
against the applied electric field $E$. 
It is noted that when 
the electron field is larger than 500\ V/cm, hot-electron effect\cite{conwell}
starts to play an important role.\cite{weng}
It is seen from the figure that the $\tau$-$E$
dependence  is similar to the $\tau$-$T$ dependence.
Figure\ 3(b) shows the strain dependence of the SDT under
different electric fields. Again, one observes a peak
under certain strain. 
These behaviors are understood as the electric field also affects the spin 
dephasing in two competing ways: On one hand, it drives the
electrons to higher momentum states;  On the other hand, it
raises the hot-electron temperature and therefore the
scattering is strengthened.

Finally, as $\langle k_\xi^2\rangle$  
depends not only on temperature, but also on electron density, 
we show  the strain
dependence of the SDT at different electron densities.
The external electric field is assumed to be zero. The 
result is summarized  in  Fig.\ 4. One finds
that the $\tau$-$\epsilon$ dependence also shows
a peak for each electron density. Moreover, the peak moves towards small
strain when the electron density increases. This is in consistent with the
fact that  $\langle k_\xi^2\rangle$ increases with the electron density.

In conclusion,  we have studied the effect of strain on the spin dephasing
in (001) GaAs QW's with a small well width under different 
conditions such as temperature, electric field and electron density. 
We show that one can effectively adjust the Dresselhaus spin
splitting via strain in two dimension case. Especially 
at certain conditions the Dresselhaus spin splitting can be
mostly canceled by the strain and one may get an {\em extremely} long
SDT (up to nanoseconds in comparison to
tens of picoseconds in ordinary strain-free sample) 
in narrow GaAs QW's.  This provides a
unique way to control the spin coherence and
get two-dimensional devices with extremely long SDT.

This work was supported by the Natural Science Foundation of China
under Grant Nos. 90303012 and 10247002, the Natural Science Foundation 
of Anhui Province under Grant No. 050460203 and SRFDP.

\end{document}